\pdfoutput=1
\documentclass[article]{jss}

\usepackage{amsmath} 
\usepackage{amsfonts} 
\usepackage{amssymb} 
\usepackage{graphicx} 
\usepackage{url} 
\usepackage{enumitem} 
\usepackage{float} 
\usepackage{booktabs} 
\usepackage{listings} 

\author{Nathan Green \\Department of Statistical Science,\\ UCL, London, UK}
\title{Population-Adjusted Indirect Treatment Comparison with the \pkg{outstandR} Package in \proglang{R}}

\Plaintitle{Indirect Treatment Comparison with the outstandR Package in R} 
\Shorttitle{\pkg{outstandR}: Indirect Treatment Comparison} 

\Abstract{
Indirect treatment comparisons (ITCs) are essential in Health Technology Assessment (HTA) when head-to-head clinical trials are absent. A common challenge arises when attempting to compare a treatment with available individual patient data (IPD) against a competitor with only reported aggregate-level data (ALD), particularly when trial populations differ in effect modifiers. While methods such as Matching-Adjusted Indirect Comparison (MAIC) exist to adjust for these cross-trial differences, they are increasingly being superseded by regression-based marginalization methods. Historically, software implementations for these methods have often been fragmented or limited in scope.

This article introduces \pkg{outstandR}, an \proglang{R} package designed to provide a comprehensive and unified framework for population-adjusted indirect comparison (PAIC). \pkg{outstandR} implements advanced G-computation methods—within both maximum likelihood and Bayesian frameworks, and Multiple Imputation Marginalization (MIM) to address non-collapsibility. By streamlining the workflow of covariate simulation, model standardization, and contrast estimation, \pkg{outstandR} enables robust and compatible evidence synthesis in complex decision-making scenarios.
}
\Plainkeywords{G-computation, simulated treatment comparison, matching-adjusted indirect comparison, bayesian, multiple imputation, R} 

\Address{
  Nathan Green\\
  Department of Statistical Science\\
  UCL\\
  E-mail: \email{n.green@ucl.ac.uk}\\
  URL: \url{https://n8thangreen.github.io/}
}

\lstdefinestyle{textttstyle}{
    basicstyle=\ttfamily, 
    breaklines=true,       
    columns=fullflexible   
}

\begin{document}

\section{Introduction}
\label{sec:introduction}

Indirect treatment comparisons (ITCs) are essential in Health Technology Assessment (HTA) when head-to-head clinical trials are absent. A common and significant challenge arises when attempting to compare a treatment with available individual patient data (IPD) against a competitor with only reported aggregate-level data (ALD). When trial populations differ in effect modifiers, indirect comparisons that do not adjust for covariates are biased; thus, population adjustment methods are increasingly used to bridge these cross-trial differences \citep{Phillippo2016DSU}.

Well-established methods for this scenario include Matching-Adjusted Indirect Comparison (MAIC)  \citep{signorovitch2010} and Simulated Treatment Comparison (STC) \citep{caro2010,ishak2015}. However, these methods often face methodological limitations, such as sensitivity to poor covariate overlap (in MAIC) or bias. Standard MAIC typically only equates the means (first marginal moment), leaving residual confounding if the variances or joint distributions remain severely mismatched. Furthermore, from a practical standpoint, software implementations for these methods have often been fragmented, limited in scope, or reliant on ad-hoc scripts.

To address methodological limitations, the literature has moved towards advanced approaches like G-computation \citep{RemiroAzocar2022a}, Multiple Imputation Marginalization (MIM) \citep{RemiroAzocar2022b} and Multilevel Network Meta-Regression (ML-NMR) \citep{phillippo_multinma_2020}. G-computation and MIM allow for model-based standardization via covariate simulation, overcoming limited patient-level data.

%

This article introduces \pkg{outstandR}, an \proglang{R} package designed to address these inefficiencies by providing a comprehensive and unified framework for population-adjusted indirect comparison (PAIC). \pkg{outstandR} streamlines the entire workflow—from covariate simulation and model standardization to contrast estimation. Beyond standard weighting (MAIC) approaches, the package implements advanced G-computation and MIM methods within both maximum likelihood and Bayesian frameworks.

The remainder of this paper is organized as follows. Section~\ref{sec:methodology} establishes the methodological framework, defining the general problem of indirect treatment comparisons and detailing the statistical mechanics of population adjustment methods, including MAIC, STC, and advanced G-computation approaches. Section~\ref{sec:implementation} describes the software architecture of \pkg{outstandR}, illustrating how the package leverages an S3 object-oriented design to separate statistical strategies from computational execution and streamline the workflow. Section~\ref{sec:examples} provides a practical demonstration of the package, guiding the user through illustrative examples using synthetic datasets for binary and continuous outcomes. Finally, Section~\ref{sec:conclusion} concludes with a discussion of the package's capabilities and potential areas for future development.

\section{Population-Adjusted Indirect Comparison (PAIC)}
\label{sec:methodology}
Consider a scenario with two trials: an $AC$ trial with individual patient data (IPD) comparing treatments $A$ and $C$, and a $BC$ trial with only published aggregate-level data (ALD) comparing treatments $B$ and $C$. The objective is to estimate a comparison of the effects of treatments $A$ and $B$ on a suitable scale in some target population $P$, denoted by the estimand $d_{AB(P)}$ \citep{RemiroAzocar2022a}. We will focus on population adjustment methods to make ``anchored'' indirect comparisons, where the evidence is connected by a common comparator, in our case treatment $C$.

\subsection{General problem formulation}
\label{ssec:problem_formulation}

Let $\mu_{A(AC)}$ and $\mu_{C(AC)}$ represent the expected outcomes for treatments $A$ and $C$ within the $AC$ trial population. We also have estimates $\hat{\mu}_{A(AC)}$ and $\hat{\mu}_{C(AC)}$ of the expected outcomes, with an equivalent set for the $BC$ trial. 

Define estimators $\hat{\Delta}_{AC(AC)}$ and $\hat{\Delta}_{BC(BC)}$ of the relative treatment effects at the trial level for a given scale. Assuming a {\it difference} scale (e.g., log-odds ratio, risk difference), the estimate of relative treatment effect between treatments $A$ and $C$
\begin{equation}
\label{eqn:relative_treatment_effect}
\hat{\Delta}_{AC(AC)} = g(\hat{\mu}_{A{(AC)}}) - g(\hat{\mu}_{C{(AC)}}),
\end{equation}
where $g(\cdot)$ is a link function (e.g., identity for linear regression, logit for logistic regression, log for Cox regression hazard ratio). An equivalent equation can be written for $\hat{\Delta}_{BC(BC)}$. If we assume that there is no difference in effect modifiers between trials, then the relative treatment is constant between trial populations. 
It follows that for any target population $P$ the estimator of the relative treatment effect $d_{AB(P)}$ is
\[
\hat{\Delta}_{AB(P)} = \hat{\Delta}_{AC(AC)} - \hat{\Delta}_{BC(BC)}.
\]
However, when distributions of the effect modifiers are different between trial populations, the relative treatment effects no longer stay constant, so they cannot simply be combined as above. This is the purpose of population-adjustment in ITC, to enable the comparison of unbiased statistics
\begin{equation}
\label{eqn:paic}
\hat{\Delta}_{AB(BC)} = \hat{\Delta}_{AC(BC)} - \hat{\Delta}_{BC(BC)}.
\end{equation}
Section~\ref{sec:paic-methods} will detail the PAIC methods available in \pkg{outstandR}.



\subsection{Population adjustment methods}
\label{sec:paic-methods}
The \pkg{outstandR} package allows the implementation of a range of methods to adjust the IPD to the target population, commonly the ALD.


\subsubsection{Matching-Adjusted Indirect Comparison (MAIC)}
Matching-Adjusted Indirect Comparison (MAIC) is a method-of-moments weighting approach, to match the aggregate characteristics of the comparator trial \citep{signorovitch2010}. The core idea of MAIC is to re-weight the patients in the IPD $AC$ trial such that their aggregated baseline characteristics match those of the ALD $BC$ trial. This re-weighting process is essentially an optimization problem.

Let $X_i$ be the vector of baseline characteristics for the $i$-th patient in the IPD $AC$ trial, for $i = 1, \dots, n_A$, where $n_A$ is the number of patients in $AC$ trial. Let $\bar{X}_B$ be the vector of mean baseline characteristics reported for the ALD $BC$ trial.

The objective is to find a set of weights, $w_i$, for each patient $i$ in the $AC$ trial, such that the weighted average of their baseline characteristics matches the aggregate baseline characteristics of the $BC$ trial. This can be formulated as minimizing a certain objective function, typically related to the sum of weights or the effective sample size, subject to matching constraints.

A common approach involves a logistic regression-based re-weighting. The weights $w_i$ are often derived from a logistic regression-based formulation, where the parameters of this model are optimized. Specifically, the weights are defined as follows:

$$ 
w_i = \exp(\beta^\top X_i),
$$
where $\beta$ is a vector of coefficients to be estimated. The optimization problem then aims to find $\beta$ such that the weighted means of the covariates in the IPD dataset match the target means from the ALD dataset. This is equivalent to solving the following set of moment-matching equations:

$$ 
\sum_{i=1}^{n_A} w_i X_i = \sum_{i=1}^{n_A} \exp(\beta^\top X_i) X_i = n_A \bar{X}_B.
$$
This system of equations can be rewritten by centring the covariates around the target means $\bar{X}_B$:

$$ 
\sum_{i=1}^{n_A} \exp(\beta^\top (X_i - \bar{X}_B)) (X_i - \bar{X}_B) = 0. 
$$
This is the first derivative with respect to $\beta$ of the following convex objective function, which is typically minimized:

$$
\min_{\beta \in \mathbb{R}^p} \sum_{i=1}^{n_A} \exp(\beta^\top (X_i - \bar{X}_B)),
$$
where $p$ is the number of baseline characteristics that are matched. 

An extension for higher-order moments is to also include the variance and the covariances. Balanced variances require the ALD to report the standard deviation, and balancing covariance requires the ALD to report the interaction term.

The minimization of this convex function ensures a unique global minimum for $\beta$. Once $\beta$ is found, the weights $w_i$ are calculated, and these weights are then used in a weighted analysis (e.g., weighted logistic regression or Cox regression) of the IPD from the $AC$ trial to estimate the treatment effect in a population adjusted to resemble the $BC$ trial.

From an information-theoretic perspective, this optimization process is mathematically equivalent to entropy balancing. It finds a set of weights that minimizes the relative entropy from the unweighted base distribution, subject to the exact constraints of matching the specified aggregate covariate moments. This ensures that the target moments are balanced, thus reducing potential bias due to differences in these specific prognostic factors or effect modifiers.

\subsubsection{Outcome Regression Models}
For an individual $i$ with outcome $Y_i$, treatment assignment $T_i \in \{0, 1\}$ (where 0 is the reference treatment and 1 is the active treatment), and vectors of baseline covariates $X_i$, the general parametric outcome model is formulated as

\begin{equation}
\label{eqn:outcome-regression}
g(\mathbb{E}[Y_i \mid X_i, T_i]) = \alpha + X_i^\top \beta_0 + \left( \beta_{trt} + X_i^\top \beta_1 \right) \mathbb{I}(T_i = 1),
\end{equation}
\noindent
where $g(\cdot)$ is the link function (e.g., identity, logit, or log), $\alpha$ is the intercept, and $\beta_0$ represents the vector of main effect coefficients for the baseline covariates. For the comparator treatment compared to the reference treatment, $\beta_{trt}$ is the conditional main effect of treatment, and $\beta_1$ is the vector of treatment-covariate interaction coefficients. If a specific covariate acts purely as a prognostic factor and does not modify the treatment effect, its corresponding parameter in $\beta_1$ is zero. Finally, $\mathbb{I}(T_i = 1)$ is an indicator function taking the value 1 if individual $i$ received treatment 1 and 0 otherwise.

\subsubsection[Simulated Treatment Comparison (STC)]{Simulated Treatment Comparison (STC)\footnote{STC is discouraged from use but included here for completeness.}}
Simulated Treatment Comparison (STC) relies on outcome regression models fitted to IPD, conditioning on covariates, then applying estimates to ALD \citep{caro2010, ishak2015}. STC involves building an outcome model from the IPD based on baseline characteristics, and subsequently evaluating this model at the mean covariate values of the ALD to approximate the marginal mean outcome.

\paragraph{Model Fitting}
Let $Y_i$ be the outcome for patient $i$ in the IPD $AC$ trial, and $X_i$ be the vector of baseline characteristics for patient $i$. An outcome regression model is fitted using the IPD $AC$ trial as in Equation~(\ref{eqn:outcome-regression}).

\paragraph{Expectation at Covariate Means}
The fitted model is then used to obtain the expected outcome of treatments $T = \{ A, C \}$ for the population of the $BC$ trial. Since only aggregate data (typically means and standard deviations) are available for the $BC$ trial's population, these aggregate values ($\bar{X}_{BC}$) are substituted into the fitted model to estimate the mean outcome.

The predicted mean outcome of treatment $T$ in the population of the $BC$ trial is calculated as
$$
\widehat{\mathbb{E}}[Y^T \mid \hat\theta, \bar{X}_{BC}] = 
g^{-1}(\hat{\alpha} + \bar{X}_{BC}^\top \hat{\beta}_0 + \left( \hat{\beta}_{trt} + \bar{X}_{BC}^\top \hat{\beta}_1 \right) \mathbb{I}(T = A))
$$
where $g^{-1}(\cdot)$ is the inverse of the link function, and $\hat\theta = \{ \hat\alpha, \hat\beta_0, \hat\beta_1, \hat\beta_{trt}\}$.

\paragraph{Limitations and modifications}
A significant limitation of standard STC arises when the link function $g(\cdot)$ is non-linear (e.g., logit or log link). In such cases, simply substituting the mean covariate values into the fitted model can lead to biased estimates. This occurs due to two distinct issues: \emph{aggregation bias}, which arises when evaluating non-linear models at aggregate covariate means rather than averaging individual-level predictions, and \emph{non-collapsibility}, an inherent property of certain effect measures (like odds ratios) where marginal and conditional effects naturally differ.

To address this, more advanced STC-type methods, particularly those incorporating actual simulation, are being developed. These methods aim to correctly estimate the marginal outcome in the target population, often by simulating individual patient profiles for the ALD trial's population based on their reported aggregate statistics and assumed covariate distributions. This approach is truly simulation-based and involves drawing from estimated joint covariate distributions and leads into the following methods based on G-computation.

Due to these severe limitations, the standard STC method has been formally deprecated in \pkg{outstandR} in favour of the more robust G-computation approaches.

\subsubsection{Parametric G-computation with Maximum Likelihood}
G-computation, also known as the G-formula, fits an outcome model to IPD using maximum likelihood, then predicts outcomes in the comparator population~\citep{RemiroAzocar2022a}. It is a method for estimating causal effects by standardizing outcomes across different treatment regimens. The conditional expectation of the outcome given treatment and covariates are modelled directly. When implemented using Maximum Likelihood Estimation (MLE), the process involves fitting a regression model for the outcome, and then using this to predict outcomes.

We consider a particular version of G-computation for ITC in \pkg{outstandR}.
The fundamental idea of G-computation is to predict outcomes under different treatments for each individual in a population, and then averaging these predicted outcomes.
We first fit the outcome regression as in Equation~(\ref{eqn:outcome-regression}). Then we wish to obtain the G-computation estimate of the marginal mean outcome under a given treatment $T$, defined as
\begin{equation}
\label{eqn:marginal-outcome}
\mathbb{E}[Y^{T}] = \sum_{x} \mathbb{E}[Y \mid T, \hat\theta, X=x] P(X=x).
\end{equation}
For continuous covariates, the sum becomes an integral. In essence, we calculate the average outcome if all members of the population followed the treatment $T$.

First, we specify and fit models for outcome and confounders using MLE. Then, the estimated parameters are used to predict outcomes under an alternative population. This is done by generating a synthetic cohort of size $N$ representing the $BC$ ALD trial population, and using the fitted outcome model to predict the outcome for each individual with IPD trial treatments (see Section~\ref{sec:synth-cohort}). We then average the predicted outcomes across all individuals to obtain the marginal mean outcome for treatment $T$ as follows.
\begin{equation}
\label{eqn:approx-marginal}
\widehat{\mathbb{E}}[Y^{T} \mid \hat\theta] = \frac{1}{N} \sum_{i=1}^N \hat{Y}_i(T, \hat\theta)
\end{equation}
Because we approximate an integral over the joint covariate distribution, the size of the synthetic pseudo-population, $N$, must be sufficiently large to minimize simulation error and ensure the sample average converges stably to the true marginal expectation.

\paragraph{Causal Effect Estimation}
Once the marginal mean outcomes are estimated for the respective treatments using the synthetic cohort, the relative causal effect is calculated. Depending on the desired scale, this is computed as the difference or ratio of the marginal means. For instance, the marginal mean difference (or risk difference) is simply $\hat{\Delta}_{AC(BC)} = \widehat{\mathbb{E}}[Y^{A}] - \widehat{\mathbb{E}}[Y^{C}]$. Specifically, because these marginal means represent the probability of the event for binary outcomes, we can obtain the marginal odds ratio as 
$$
\widehat{OR}_{AC} = \frac{\widehat{\mathbb{E}}[Y^{A}] / \left(1 - \widehat{\mathbb{E}}[Y^{A}]\right)}{\widehat{\mathbb{E}}[Y^{C}] / \left(1 - \widehat{\mathbb{E}}[Y^{C}]\right)}.
$$

\subsubsection{Parametric G-computation with Bayesian Inference}
G-computation can also be implemented using Bayesian inference. This is similar to the maximum likelihood version but offers several advantages over the frequentist MLE approach, particularly in quantifying and coherently propagating uncertainty \citep{RemiroAzocar2022a}, and incorporating prior knowledge. Instead of estimating point estimates for model parameters that maximize a likelihood function, the Bayesian approach estimates the full posterior distribution of these parameters. This allows for direct probability statements about the parameters and the resulting causal effects.

As for the MLE approach, we wish to estimate the marginal mean outcome under a specific treatment $T$. The core idea is to average predicted outcomes under counterfactual treatment scenarios. The Bayesian implementation of G-computation typically follows these steps:

\paragraph{Step 1: Specify and Fit Models with Bayesian Inference}

\begin{enumerate}
    \item Model Specification: Define the outcome regression models as in the frequentist approach for the outcome in Equation~(\ref{eqn:outcome-regression}).

    \item Prior Specification: Assign prior distributions to all model parameters $\theta$. These priors reflect existing knowledge or assumptions about the parameters before observing the data.
    \item Bayesian Model Fitting and Posterior Sampling:
    Using computational methods like Markov Chain Monte Carlo (MCMC) (e.g., Gibbs sampling, Metropolis-Hastings, Hamiltonian Monte Carlo), samples are drawn from the joint posterior distribution of all model parameters. The result is a set of $M$ posterior samples for parameters, $\theta^{(m)}$, for $m=1, \dots, M$.
\end{enumerate}

\paragraph{Step 2: Prediction/Simulation Step (Posterior Predictive Simulation)}
First, generate a synthetic target cohort (of size $N$) that reflects the target ALD population, using the exact same Gaussian copula approach described for the Maximum Likelihood implementation. Then, for each set of sampled parameters from the posterior distribution $(m=1, ..., M)$:
\begin{enumerate}
    \item Predict Counterfactual Outcomes: For each individual $i$ in the synthetic cohort, set their treatment to the hypothesized treatment $T$. Using the $m$-th sample of parameters for the outcome model $\theta^{(m)}$, predict the outcome for each individual based on their simulated baseline covariates.
    
    \item Calculate marginal mean for this sample: Average the predicted outcomes across all $N$ individuals in the synthetic cohort to obtain the marginal mean outcome for the treatment $T$ for this $m$-th posterior sample:
\begin{equation}
\label{eqn:marginal-outcome-bayesian}
\mathbb{E}[Y^{T} \mid \theta^{(m)}] = \sum_{x} \mathbb{E}[Y \mid T, \theta^{(m)}, X=x] P(X=x), \quad m = 1, \ldots, M;
\end{equation}
which is computed with the approximation
\begin{equation}
\label{eqn:approx-marginal-bayes}
\widehat{\mathbb{E}}[Y^{T} \mid \theta^{(m)}] = \frac{1}{N} \sum_{i=1}^N \hat{Y}_i(T, \theta^{(m)}), \quad m = 1, \ldots, M.
\end{equation}
\end{enumerate}

This results in $M$ samples of the marginal mean outcome, which together form the posterior distribution of the marginal mean under treatment $T$.

\paragraph{Step 3: Causal Effect Estimation}
To estimate the causal effect of two treatments $A$ and $C$, the above process is repeated to obtain $M$ posterior samples for each $\widehat{\mathbb{E}}[Y^{T} \mid \theta^{(m)}]$. The posterior distribution of the causal effect is then directly obtained by computing the difference for each posterior sample
$$
\hat{\Delta}^{(m)}_{AC(BC)} = \widehat{\mathbb{E}}[Y^{A} \mid \theta^{(m)}] - \widehat{\mathbb{E}}[Y^{C} \mid \theta^{(m)}]
$$
From this posterior distribution of the causal effect, one can derive point estimates, credible intervals, and posterior probabilities.

A key advantage of Bayesian G-computation is that it provides a cohesive framework for full uncertainty quantification. Specific benefits include:
\begin{itemize}
    \item Comprehensive uncertainty assessment: The posterior distributions for all parameters and causal effects natively propagate error throughout the entire estimation process.
    \item Incorporation of prior knowledge: The framework allows for the inclusion of expert knowledge or evidence from previous studies through informative prior distributions.
    \item Handling complex models: MCMC methods are highly flexible and well-suited for fitting the complex hierarchical or non-linear models often encountered in causal inference.
    \item Direct probabilistic statements: It allows analysts to make intuitive claims, such as ``there is a 95\% probability that the true causal effect falls between $a$ and $b$''.
\end{itemize}

\subsubsection{Multiple Imputation Marginalization (MIM)}
Multiple Imputation Marginalization (MIM) is an alternative framework for performing G-computation to recover a compatible marginal treatment effect. Rather than viewing standardisation purely as a prediction exercise, MIM conceptualizes the unobserved potential outcomes of the target population as a missing data problem. By leveraging the principles of multiple imputation, it provides a rigorous mechanism for standardising effects and propagating uncertainty, which is particularly useful for non-collapsible effect measures derived from generalized linear models.

In \pkg{outstandR}, MIM is implemented exclusively within a Bayesian statistical framework to seamlessly connect model fitting with the imputation of the target population's counterfactuals.
The first step is to fit an outcome regression in the same way as for the standard Bayesian G-computation above.
The next step is where MIM diverges from the previous approach. Instead of predicting a single expected value, by marginalising over a sufficiently large synthetic cohort, we ``impute'' the missing counterfactual outcomes for multiple synthetic target cohorts. For $m=1, ..., M$ draws from the posterior distribution of the outcome model parameters, $\theta^{(m)}$.

\begin{itemize}
    \item \emph{Counterfactual Prediction:} For each individual $i$ in the synthetic BC cohort, use $\theta^{(m)}$ to impute their probability of the outcome under both treatments.
    \item \emph{Marginalization within Imputations:} For each of the $M$ imputation sets, calculate the marginal expected outcome by averaging the imputed probabilities across all $N$ individuals in the synthetic cohort as in Equation~(\ref{eqn:approx-marginal-bayes}). 
    \item \emph{Calculate Effect Measures:} Compute the desired marginal effect measure (e.g., the marginal odds ratio, denoted $Q^{(m)}$) and its associated within-imputation variance ($U^{(m)}$) for each $m$.
    \item \emph{Pooling Results:} Finally, the $M$ estimates are pooled. However, because MIM simulates the target population outcomes rather than treating baseline covariates as missing, standard Rubin's rules are modified. The pooled point estimate is the average of the imputed estimates:
   $\overline{Q} = \frac{1}{M}\sum_{m=1}^{M}Q^{(m)}$
   Crucially, to calculate the total variance for this specific marginalization approach, the within-imputation variance ($\overline{U}$) is subtracted from the between-imputation variance ($B$):
   $\text{Var}(\overline{Q}) = B - \overline{U}$
   where 
   \begin{equation}
    \label{eqn:pooling-results}
   B = \frac{1}{M-1}\sum_{m=1}^{M}(Q^{(m)} - \overline{Q})^2 \quad \text{and} \quad \overline{U} = \frac{1}{M}\sum_{m=1}^{M}U^{(m)}.
   \end{equation}
\end{itemize}

\subsection{Synthetic cohort generation}\label{sec:synth-cohort}
Generally, there are several different approaches to generating a synthetic cohort. We will detail how to create one that accurately reflects the target population's covariate structure by employing a \emph{Gaussian copula}. This allows us to decouple the dependence structure (correlation) from the marginal distributions of the individual covariates.

Let $K$ be the number of covariates and $\boldsymbol{\Sigma}$ be the $K \times K$ correlation matrix, typically estimated from IPD. We wish to simulate $N$ individuals for the pseudo-population.

The simulation proceeds in three steps:

\begin{enumerate}
    \item \textbf{Generate Latent Multivariate Normal Data:}
    We first draw $N$ samples from a multivariate normal distribution with mean vector $\mathbf{0}$ and correlation matrix $\boldsymbol{\Sigma}$
    \begin{equation*}
        \mathbf{Z}_i \sim MVN_K(\mathbf{0}, \boldsymbol{\Sigma}), \quad \text{for } i = 1, \dots, N.
    \end{equation*}

    \item \textbf{Probability Integral Transform:}
    We transform these correlated normal variates into uniform variates $\mathbf{U}_i = (U_{i1}, \dots, U_{iK})$ using the cumulative distribution function (CDF) of the standard normal distribution, $\Phi(\cdot)$
    \begin{equation*}
        U_{ik} = \Phi(Z_{ik}),
    \end{equation*}
    where $U_{ik} \sim \text{Uniform}(0, 1)$ and the correlation structure of $\mathbf{Z}$ is preserved.

    \item \textbf{Inverse Probability Transform (Marginals):}
    Finally, we map the uniform variates to the target scales using the inverse CDFs (quantile functions) of the marginal distributions. Let $F_k(\cdot; \theta_k)$ denote the CDF for the $k$-th covariate with parameters $\theta_k$
    \begin{equation}
        \label{eqn:marginal-transformation}
        X_{ik} = F_k^{-1}(U_{ik}; \theta_k).
    \end{equation}
\end{enumerate}

This procedure ensures that the resulting synthetic covariates $\mathbf{X}$ possess the desired marginal properties defined by $F_k$ while maintaining the correlation structure $\boldsymbol{\Sigma}$.

A limitation of using the copula approach is that, at the moment, covariate simulation is flawed whenever non-continuous covariates are included. This occurs because the Probability Integral Transform strictly requires continuous marginals to produce a uniform distribution. When a covariate is discrete, its cumulative distribution function is a step function, resulting in a non-uniform, discrete mapping leading to  distorted correlations in the simulated synthetic cohort.

\subsection{Variance estimation and inference}
\pkg{outstandR} implements three distinct approaches for variance estimation and inference, depending on the chosen strategy and user preference: resampling-based, model-based, and robust analytical estimators.

\subsubsection{Resampling and model-based approaches}

By default, the package employs computational methods to estimate uncertainty, reflecting the complexity of population-adjusted estimators where simple analytical variance formulas are often intractable.

\begin{itemize}
    \item \textbf{Non-parametric Bootstrapping:} For frequentist strategies such as MAIC and G-computation (ML), standard errors and confidence intervals are derived via non-parametric bootstrapping \citep{efron1979bootstrap}. The IPD is resampled $R$ times, and the entire estimation procedure is repeated for each resample. The variance of the treatment effect is estimated as the sample variance of the $R$ bootstrap replicates.
    
    \item \textbf{Bayesian Posterior Inference (Bayesian G-computation only):} For the Bayesian G-computation strategy, inference is fully Bayesian. Uncertainty is quantified using the posterior distribution of the marginal treatment effect, which naturally propagates uncertainty from the outcome model parameters and the missing potential outcomes.
    
    \item \textbf{Modified Pooling Rules (MIM):} For the Multiple Imputation Marginalization (MIM) strategy, variance is estimated using modified pooling rules. Because MIM simulates the target population outcomes rather than treating baseline covariates as missing, standard Rubin's rules do not apply. Instead, the total variance is calculated by subtracting the within-imputation variance from the between-imputation variance, given in Equation~(\ref{eqn:pooling-results}) \citep{rubin1987}.
\end{itemize}

\subsubsection{Robust sandwich estimator} \label{sec:variance_methods}
As an alternative to computationally intensive resampling, \pkg{outstandR} also implements a robust sandwich variance estimator for the MAIC. This provides a computationally efficient way to obtain standard errors that are robust to model misspecification (e.g., heteroscedasticity) when performing weighted analyses.

For a marginal treatment effect estimate $\hat{\Delta}$, which is a function $g(\cdot)$ of the estimated model parameters $\hat{\boldsymbol{\theta}}$, the variance is approximated using the delta method:
\begin{equation}
\widehat{\text{Var}}(\hat{\Delta}) = \nabla g(\hat{\boldsymbol{\theta}})^\top \cdot \mathbf{V}_{\text{sand}} \cdot \nabla g(\hat{\boldsymbol{\theta}}),
\end{equation}
where $\mathbf{V}_{\text{sand}}$ is the heteroscedasticity-consistent (HC) covariance matrix of the regression coefficients $\hat{\boldsymbol{\theta}}$, computed as $\mathbf{H}^{-1} \mathbf{M} \mathbf{H}^{-1}$ where $\mathbf{H}$ is the ``bread'' (Hessian/Fisher Information) and $\mathbf{M}$ is the ``meat'' (outer product of gradients) of the likelihood function. The gradient $\nabla g(\hat{\boldsymbol{\theta}})$ captures how the marginal effect changes with respect to the underlying model parameters and is computed numerically to support complex standardization steps.

\section{Implementation and software design}
\label{sec:implementation}

The \pkg{outstandR} package is implemented in \proglang{R} and designed for ease of use while providing powerful statistical capabilities. The core functionality is encapsulated in the main \texttt{outstandR()} function.

\subsection{The outstandR() function} \label{ssec:outstandr_function}

The primary function \texttt{outstandR()} orchestrates the standardisation process. Its signature is:

\begin{itemize}[noitemsep,topsep=0pt]
    \item \texttt{ipd\_trial}: Individual-level patient data (e.g., for treatments $A$ and $C$), expected in a long format including treatment and outcome columns consistent with the formula object.
    \item \texttt{ald\_trial}: Aggregate-level data (e.g., for treatments $B$ and $C$). This should contain columns for \code{variable} (covariate name), \texttt{statistic} (e.g., \code{mean}, \code{sd}, \code{prop}, \code{sum}), \code{value} (numerical summary statistic), and \code{trt} (treatment label, or \texttt{NA} if common covariate distribution is assumed).
    \item \texttt{strategy}: A computation strategy function, which can be one of \texttt{strategy\_maic()}, \texttt{strategy\_stc()}, \texttt{strategy\_gcomp\_ml()}, \texttt{strategy\_gcomp\_bayes()}, or \texttt{strategy\_mim()}.
    \item \texttt{ref\_trt}: The reference, common, or anchoring treatment name. Note that this can also be called the {\it index} and the comparator called the reference.
    \item \texttt{CI}: The desired confidence interval level (between 0 and 1, default 0.95).
    \item \texttt{scale}: The relative treatment effect scale. If \texttt{NULL}, it is automatically determined from the model. Options include \code{log_odds}, \code{log\_relative\_risk}, \code{risk\_difference}, \code{mean\_difference}, \code{rate\_difference} depending on data type.
    \item \texttt{var\_method}: Method of variance estimation. Options are \code{sample} (default) and \code{sandwich}.
    \item \texttt{verbose}: Logical (default \texttt{TRUE}). Controls console output and provides proactive warnings for computationally expensive operations (e.g., high \texttt{n\_boot} or large pseudo-populations) to prevent users from assuming the session has hung.
    \item \texttt{seed}: Random number seed (optional).
    \item \texttt{...}: Additional arguments, allowing named arguments to be passed to \texttt{rstanarm::stan\_glm()} via \texttt{strategy\_gcomp\_bayes()}.
\end{itemize}

\begin{table}[ht!]
\centering
\caption{Structure of the \texttt{outstandR} S3 class object. The object acts as a container for both relative contrasts and absolute predictions, along with metadata attributes describing the analysis settings.}
\label{tab:s3_structure}
\begin{tabular}{l l p{9.5cm}}
\toprule
\textbf{Element} & \textbf{Class} & \textbf{Description} \\
\midrule
\texttt{results} & List & Container for the main estimation statistics. \\
\hspace{3mm} \texttt{\$contrasts} & List & Container for relative treatment effect statistics. \\
\hspace{6mm} \texttt{\$means} & Numeric & Point estimates for pairwise contrasts (e.g., AB, AC, BC). \\
\hspace{6mm} \texttt{\$variances} & Numeric & Variance estimates corresponding to the contrasts. \\
\hspace{6mm} \texttt{\$CI} & List & Confidence intervals (lower and upper bounds) for the contrasts. \\
\hspace{3mm} \texttt{\$absolute} & List & Container for absolute (arm-specific) estimates. \\
\hspace{6mm} \texttt{\$means} & Numeric & Adjusted mean outcomes for specific treatments. \\
\hspace{6mm} \texttt{\$variances} & Numeric & Variances of the adjusted absolute means. \\
\hspace{6mm} \texttt{\$CI} & List & Confidence intervals for absolute means. \\
\addlinespace
\texttt{call} & Language & Original function call. \\
\texttt{outcome\_model} & Character & The covariates, treatments and outcome formula. \\
\texttt{balance\_model} & Character & (MAIC) A covariates formula with empty left hand side. \\
\texttt{scale} & Character & The scale of the treatment effect (e.g., \texttt{log\_odds}, \texttt{risk\_difference}). \\
\texttt{family} & Character & The statistical model family (e.g., \texttt{binomial}, \texttt{gaussian} or \texttt{poisson}). \\
\texttt{ref\_trt} & Character & Name of the common comparator or anchoring treatment. \\
\texttt{ipd\_comp} & Character & Name of the IPD study-specific comparator. \\
\texttt{ald\_comp} & Character & Name of the aggregate level data study comparator. \\
\texttt{CI} & Numeric & The confidence level used for intervals (e.g., 0.95). \\
\texttt{var\_method} & Character & Method used for variance estimation (\texttt{sample}, \texttt{sandwich} or \texttt{pool}). \\
\texttt{model} & List & Container for method-specific elements. \\
\hspace{3mm} \texttt{\$method\_name} & Character & From \code{MAIC}, \code{STC}, \code{GCOMP\_BAYES}, \code{GCOMP\_ML}, and \code{MIM}. \\
\hspace{3mm} \texttt{\$rho} & Numeric & (G-computation, MIM) Matrix of correlation coefficients. \\
\hspace{3mm} \texttt{\$N} & Numeric & (G-computation, MIM) Synthetic sample size for the pseudo-population. \\
\hspace{3mm} \texttt{\$n\_boot} & Numeric & (MAIC, G-computation ML) Bootstrap replication size. \\
\hspace{3mm} \texttt{\$weights} & Numeric & (MAIC) Estimated weights for IPD individuals. \\
\hspace{3mm} \texttt{\$ESS} & Numeric & (MAIC) Effective Sample Size. \\
\hspace{3mm} \texttt{\$fit} & Object & (STC, G-computation) The fitted model object (e.g., \texttt{glm} or \texttt{stanfit}). \\
\hspace{3mm} \texttt{\$stan\_args} & List & (Bayesian G-computation, MIM) Arguments passed to the Stan engine. \\
\hspace{3mm} \texttt{\$n\_imp, \$nu, \$hats.v} & Numeric & (MIM) Multiple imputation statistics. \\
\hspace{3mm} \texttt{\$marginal\_params} & List & (G-computation) Marginal distributions parameters. \\ 
\hspace{3mm} \texttt{\$marginal\_distns} & Vector & (G-computation) Marginal distributions names. \\ 
\bottomrule
\end{tabular}
\end{table}

The function returns a list of statistics as an \texttt{outstandR} S3 class object. The output elements are shown in Table~\ref{tab:s3_structure}. Results are presented as both contrasts and absolute values. For each of these, the means, variances and confidence intervals are provided. Additional elements include the confidence level, reference treatment name, outcome scale, roles of the different treatments, variance estimation method and type of likelihood model. These can be thought of as the arguments provided to the call. A \texttt{model} list is also included with ITC algorithm-specific elements. These are useful for performing diagnostics and more detailed model interrogation.

Although the output to the console and Table 1 refer broadly to a CI, users should interpret this as a frequentist confidence interval for MAIC and G-computation (ML), and as a Bayesian credible interval for the Bayesian G-computation and MIM strategies.

\subsubsection{How the outstandR() function works internally}
The \texttt{outstandR()} function is a high-level wrapper function to carry out all required steps of ITC and population adjustment depending on the chosen method. It is designed to abstract away the task of transporting from the IPD to the ALD population and computing the contrast and absolute statistics of interest. The commonality of code across methods is exploited to simplify the code and adhere to the DRY principle. 

\subsection{The ``strategy'' family of functions}
\pkg{outstandR} provides a set of functions called \emph{strategies} that define the type of approach to use. The generic signature is

\begin{verbatim}
strategy_*(formula, family, trt_var, ...)
\end{verbatim}

\begin{itemize}
\item \texttt{formula}: A list explicitly separating the models, containing \texttt{outcome\_model} and/or \texttt{balance\_model} formulae. 
    \begin{itemize}
        \item \texttt{outcome\_model}: A regression formula for predicting the outcome, containing prognostic factors (PF) and effect modifiers (EM) (e.g., \texttt{y $\sim$ pf + trt + trt:em}).
        \item \texttt{balance\_model}: A one-sided formula that defines covariates to balance (e.g., \texttt{$\sim$ x1 + x2}).
    \end{itemize}
    To streamline the user experience, \pkg{outstandR} features intelligent defaults. If \texttt{balance\_model} is omitted, it defaults to a linear sum of all covariates found in the aggregate-level data. If a traditional single formula is passed (legacy support), the package automatically infers the \texttt{balance\_model} by stripping out the response and treatment variables.
    \item \texttt{family}: A base \proglang{R} \pkg{stats} package \code{family} object specifying the distribution and link function (e.g., \code{binomial}). Family objects provide a convenient way to specify the details of the models used by functions such as \texttt{glm()}. See \texttt{stats::family()} for more details.
    \item \texttt{trt\_var}: Treatment variable name; string e.g. \texttt{trt}. If not provided then heuristics are used to guess from the \texttt{outcome\_model} formula.
\end{itemize}

The asterisk indicates a placeholder for the name of the particular approach from those detailed in the Methods in Section~\ref{sec:methodology}. Table~\ref{tab:strategy_options} gives the strategy functions and optional arguments of their signature, represented by the ellipses.
\begin{table}[htbp]
    \centering
    \caption{Strategy functions corresponding to each approach and optional arguments available in \pkg{outstandR}. $^{\dagger}$STC is deprecated as of v1.0.1}
    \label{tab:strategy_options}
    \begin{tabular}{lll}
        \toprule
        \textbf{Method} & \textbf{Strategy Function} & \textbf{Optional Arguments} \\
        \midrule
        Matching-Adjusted Indirect Comparison & \texttt{strategy\_maic()} & \texttt{n\_boot} \\
                           &                            & \texttt{moments} \\
                           &                            & \texttt{int} \\
        \addlinespace
        Simulated Treatment Comparison$^{\dagger}$ & \texttt{strategy\_stc()} & \texttt{N} \\
        \addlinespace
        G-Computation (ML) & \texttt{strategy\_gcomp\_ml()} & \texttt{rho} \\
                           &                            & \texttt{marginal\_distns} \\
                           &                            & \texttt{marginal\_params} \\
                           &                            & \texttt{N} \\
                           &                            & \texttt{n\_boot} \\
        \addlinespace
        G-Computation (Bayesian) & \texttt{strategy\_gcomp\_bayes()} & \texttt{rho} \\
                                 &                                & \texttt{marginal\_distns} \\
                                 &                                & \texttt{marginal\_params} \\
                                 &                                & \texttt{N} \\
        \addlinespace
        Multiple Imputation Marginalization & \texttt{strategy\_mim()} & \texttt{rho} \\
                           &                            & \texttt{marginal\_distns} \\
                           &                            & \texttt{marginal\_params} \\
            &                        & \texttt{N} \\
        \bottomrule
    \end{tabular}
\end{table}
The argument definitions are:

\begin{itemize}
    \item \texttt{rho}: A named square matrix of covariate correlations; default \texttt{NA}.
    \item \texttt{marginal\_distns}: Marginal distributions names; vector default \texttt{NA}.
    \item \texttt{marginal\_params}: Marginal distributions parameters; list of lists, default \texttt{NA}.
    \item \texttt{n\_boot}: Number of resamples used in non-parametric bootstrap; default 1000.
    \item \texttt{N}: Synthetic population size for G-computation. Ensure sufficiently large to minimize simulation error and that the sample average converges stably to the true marginal expectation; default 1000.
    \item \texttt{moments}: Integer specifying the highest moment to balance; default 1. Setting to 2 automatically derives the squared term target using the mean and SD from the ALD.
    \item \texttt{int}: Logical; default FALSE. If TRUE, expands the design matrix to include two-way interactions to balance covariances.
\end{itemize}
The resulting objects from running a strategy-type function each have their own \texttt{print} methods.

\subsection{Related R packages}
Several \proglang{R} packages share conceptual ground with the methods implemented here, although they typically differ in scope or intended application.

General-purpose tools for marginal effects and standardization exist, but often focus on single-study contexts. For example, the \pkg{marginaleffects} package \citep{aari_marginaleffects_2024} offers a comprehensive suite for post-estimation quantities but is not explicitly designed for cross-study population adjustment. Similarly, \pkg{stdReg2} focuses on standardizing outcomes within a single dataset \citep{g_sofer_2023_10022204}.

For propensity score weighting and covariate balance assessment, packages such as \pkg{WeightIt} \citep{greifer_weightit_2024} and \pkg{cobalt} \citep{greifer_cobalt_2024} provide powerful, general-purpose infrastructure. \pkg{WeightIt} offers unified interfaces for generating balancing weights for various estimands, including the ability to balance higher-order moments. Also, \pkg{cobalt} provides extensive tools for diagnosing covariate balance before and after adjustment. However, like the marginalization tools, they are primarily built for single-study causal inference rather than the specific transportability challenge of synthesizing individual patient data with aggregate-level data targets.

In the domain of longitudinal data, \pkg{gfoRmula} \citep{sjolander2023gformula} and \pkg{gFormulaMI} \citep{Sterne2023} implement the G-formula to estimate causal effects in the presence of time-varying treatments and confounders. However, these are primarily designed for longitudinal analysis of a single study (the latter incorporating multiple imputation via \pkg{mice}) rather than the synthesis of disparate data sources.

Closer to the context of ITC, the \pkg{multinma} package \citep{phillippo_multinma_2020} implements multilevel network meta-regression to synthesize individual and aggregate data. It addresses population adjustment within a Bayesian framework using numerical integration, distinct from the modular, frequentist G-computation or weighting approaches offered here. Finally, \pkg{maicplus} \citep{maicplus_2024} is a specialist ITC package, but its functionality is strictly limited to the Matching-Adjusted Indirect Comparison (MAIC) methodology.

\subsection{Software architecture and S3 class design}
We adopted a functional object-oriented design using \proglang{R}'s S3 system to balance modularity with ease of use. The core design philosophy separates the \emph{definition} of the statistical approach from its \emph{execution}.
This is achieved through \code{strategy} objects. Functions such as \code{strategy_maic()}, \code{strategy_stc()}, and \code{strategy_gcomp_ml()} do not perform any immediate calculation. Instead, they validate the user's hyperparameters (e.g., formula, error distribution, bootstrap iterations) and return a structured list object with a specific class attribute (e.g., \code{c("maic", "strategy", "list")}).

The main wrapper function \code{outstandR()} remains agnostic to the specific method being used, providing a unified interface. It accepts a \code{strategy} object and delegates the computational heavy lifting to the appropriate backend. This simplifies the user API, as the function signature for \code{outstandR()} remains consistent regardless of whether the user is performing a simple MAIC or a complex Bayesian G-computation. Also, new methods can be added by simply defining a new constructor function (e.g., \code{strategy_newmethod()}) and a corresponding S3 method for the internal generic \code{calc_IPD_stats()}. This allows the package to grow without modifying the core orchestration logic.

The \pkg{outstandR} package utilizes \proglang{R}'s S3 dispatch system to provide a default \texttt{plot()} method. This method automatically extracts both relative contrasts and absolute arm-specific estimates from the results object. Furthermore, the method is designed to accept multiple \texttt{outstandR} objects, enabling direct visual comparison of different adjustment strategies or sensitivity analyses within a single, faceted forest plot.

\subsection{Internal method dispatch}
The \pkg{outstandR} package relies on standard S3 method dispatch to route the analysis to the correct computational engine. The internal generic function \code{calc_IPD_stats()} is the central dispatch point.
When \code{outstandR()} is called, it pre-processes the input data and then calls \code{calc_IPD_stats(strategy, analysis_params)}. \proglang{R}'s S3 system inspects the class of the \code{strategy} object and dispatches to the specific implementation:

\begin{itemize}
  \item \code{calc_IPD_stats.maic()}: Handles moment-matching weighting and bootstrap estimation.
  \item \code{calc_IPD_stats.stc()}: Manages outcome regression modelling and prediction.
  \item \code{calc_IPD_stats.gcomp_ml()}: Orchestrates the simulation of pseudo-populations and maximum likelihood estimation.
  \item \code{calc_IPD_stats.gcomp_bayes()}: Interfaces with the \proglang{Stan} backend for Bayesian inference.
  \item \code{calc_IPD_stats.mim()}: Interfaces with the \proglang{Stan} backend for multiple imputation and Bayesian inference.
\end{itemize}

To adhere to the DRY (Don't Repeat Yourself) principle, we utilized a functional programming pattern with the \code{IPD_stat_factory()} helper. Since many strategies require similar post-processing---such as calculating the Average Treatment Effect (ATE) from arm-specific means and formatting the output list---\code{IPD_stat_factory()} wraps the specific calculation functions. It standardizes the output structure (means, variances, confidence intervals) across all methods, ensuring that the \code{outstandR} return object is consistent and predictable for the user.

\section{Using the outstandR package}
\label{sec:examples}
This section gives a practical demonstration of the core functionality available in the package.

\subsection{Data requirements and structure} \label{sec:data-req}
The \pkg{outstandR} package requires two primary inputs: individual patient data (IPD) and aggregate-level data (ALD). While the statistical content varies by outcome type, the structural format is standardized across the package to ensure a consistent workflow.

\subsubsection{Individual patient data (IPD)}
IPD must be provided in long format, where each row represents a single patient. It must contain the following columns:
\begin{itemize}
    \item Outcome: A column representing the observed result (e.g., 0/1 for binary, numeric for continuous, or counts). The name is user-specified but defaults to \texttt{y}.
    \item Treatment: A character or factor column identifying the treatment arm (e.g., A or C). The name is user-specified but defaults to \texttt{trt}.
    \item Covariates: Separate columns for each prognostic factor (PF) and effect modifier (EM) used in the adjustment.
\end{itemize}

\subsubsection{Aggregate-level data (ALD)}
ALD must be provided in a ``tidy'' long format to facilitate mapping to IPD covariates. The ALD data frame requires four specific columns:
\begin{itemize}
    \item \texttt{variable}: The name of the covariate or outcome (must match IPD column names).
    \item \texttt{statistic}: The type of summary measure provided (e.g., \code{mean}, \code{sd}, \code{prop}, \code{sum}, or \code{N}).
    \item \texttt{value}: The numerical value of the statistic.
    \item \texttt{trt}: The treatment label (e.g., B or C). For baseline covariates assumed to be balanced, this may be \texttt{NA}.
\end{itemize}

Table~\ref{tab:outcome_data_summary} shows the differences in the IPD and ALD data frames for the difference outcome types.

\begin{table}[h!]
\centering
\caption{Summary of outcome data types and required aggregate-level statistics.}
\label{tab:outcome_data_summary}
\begin{tabular}{lll}
\hline
\textbf{Outcome Type} & \textbf{IPD Outcome ($y$)} & \textbf{ALD Outcome Statistics} \\ \hline
Binary & Integer (0, 1) & \texttt{mean}, \texttt{sum}, \texttt{N} \\
Continuous & Numeric & \texttt{mean}, \texttt{sd}, \texttt{N} \\
Count & Integer ($\ge 0$) & \texttt{mean}, \texttt{sum}, \texttt{N} \\ \hline
\end{tabular}
\end{table}

\subsubsection{Available internal data sets}
To illustrate the use of \pkg{outstandR}, we show its application with synthetic datasets. These are contained in the \pkg{outstandR} package and can be loaded via the \texttt{data()} function. The naming convention adopted is \newline \\
\texttt{<comparator><reference>\_<level>\_<outcome>Y\_<covariates>X} \newline
\\
where comparator and reference are the treatment arm names, level is either \texttt{IPD} (individual-level patient data) or \texttt{ALD} (aggregate-level data), outcome is \texttt{bin} (binary), \texttt{count} (count) or \texttt{cont} (continuous), and covariates is \texttt{bin}, \texttt{cont} or \texttt{mixed}. For example, \texttt{AC\_IPD\_binY\_binX} is the name of the data set for treatments $A$ vs $C$, patient-level data, and both binary-valued outcome and covariates.

\subsection{Installation}
First, install \pkg{outstandR} from CRAN: 

\begin{verbatim}
> install.packages("outstandR")
\end{verbatim}
or the development version from GitHub using R-universe:
\begin{verbatim}
> install.packages("outstandR",
+                  repos = c("https://statisticshealtheconomics.r-universe.dev",
+                  "https://cloud.r-project.org"))
\end{verbatim}

\subsection{Illustrative examples}
Attach the package to the current environment.

\begin{verbatim}
> library(outstandR)
\end{verbatim}

\subsubsection{Binary outcome data}
In the first example, consider binary outcome data and continuous covariates only.
Load the aggregate level and individual-patient data into the environment.

\begin{verbatim}
> data(AC_IPD_binY_contX) # A vs C individual-level patient data
> data(BC_ALD_binY_contX) # B vs C aggregate-level data
\end{verbatim}

Inspecting the IPD and recalling Section~\ref{sec:data-req}, we can see that each row corresponds to an individual and the first four columns correspond to covariates, two effect modifiers only, and two prognostic factors only. These are all continuous valued and the outcome is binary valued 0, 1.

\begin{verbatim}
> # individual-level data
> head(AC_IPD_binY_contX)

  PF_cont_1   PF_cont_2   EM_cont_1   EM_cont_2 trt y    
1 0.0826453  0.71645068  0.66422255 -0.09202652   A 0 
2 0.6957433  0.38608723  0.79875530  0.74880584   C 1  
3 0.6516889  0.50572444  0.39944706 -0.07230380   A 0 
4 0.2165975  0.35783123  0.25957984 -0.55405590   A 0 
5 0.2605580 -0.05826527 -0.10696235  0.33762456   A 0 
6 0.1110818  0.64041589  0.04867791  0.34980415   C 1 
\end{verbatim}

Similarly, the ALD has the corresponding covariates but not as separate columns (wide format) but in a long format with all covariate names in the \texttt{variable} column.  Notice that the treatment column \texttt{trt} is NA for the covariates because we assume that the covariate distribution is balanced to be the same between treatment arms.

\begin{verbatim}
> # aggregate-level data
> BC_ALD_binY_contX

# A tibble: 16 × 4
   variable  statistic   value trt  
   <chr>     <chr>       <dbl> <chr>
 1 EM_cont_1 mean        0.651 <NA> 
 2 EM_cont_1 sd          0.391 <NA> 
 3 EM_cont_2 mean        0.592 <NA> 
 4 EM_cont_2 sd          0.416 <NA> 
 5 PF_cont_1 mean        0.653 <NA> 
 6 PF_cont_1 sd          0.371 <NA> 
 7 PF_cont_2 mean        0.583 <NA> 
 8 PF_cont_2 sd          0.437 <NA> 
 9 y         mean        0.615 C    
10 y         sd          0.490 C    
11 y         sum        40     C    
12 y         mean        0.274 B    
13 y         sd          0.448 B    
14 y         sum        37     B    
15 <NA>      N          65     C    
16 <NA>      N         135     B
\end{verbatim}

The next step is to define the formula, which specifies the relationship between the covariates, treatments and outcome.
For this example, the outcome regression is

$$
\text{logit}(p) = \alpha + \beta_{01} X_{1} + \beta_{02}X_{2} + (\beta_{trt} + \beta_{13} X_{3} + \beta_{14} X_{4}) \mathbb{I}(trt = 1),
$$
The corresponding \proglang{R} formula used by \pkg{outstandR} is independent of the particular type of data and so we do not include a link function at this stage, i.e, the identity link.
\begin{verbatim}
> # linear formula for the outcome model
> lin_form_contX <- as.formula(
+    "y ~ PF_cont_1 + PF_cont_2 + trt + trt:EM_cont_1 + trt:EM_cont_2")
\end{verbatim}
That is, the main effects are the prognostic factors (\texttt{PF\_cont\_1} and \texttt{PF\_cont\_2}) and the treatment  \texttt{trt}. The interaction terms are with the treatment defined as effect modifiers (\texttt{EM\_cont\_1} and \texttt{EM\_cont\_2}). 

For the MAIC approach, we define the balancing formula to include the effect modifiers as follows 
\begin{verbatim}
> # linear formula for the outcome model
> bal_form_contX <- as.formula("~ EM_cont_1 + EM_cont_2")
\end{verbatim}

\paragraph{Relative treatment effects}
Let $p_C$ be the proportion of events in the comparator treatment arm and $p_B$ be the proportion of events in the reference treatment arm. For the logit link with binary data, the mean treatment effect required in Equation~(\ref{eqn:paic}) is the log-odds ratio
$$
\hat{\Delta}_{BC} = \text{logit}(p_B) - \text{logit}(p_C)
$$
with variance
$$
\text{Var}(\Delta_{BC}) = \frac{1}{y_B} + \frac{1}{N_B - y_B} + \frac{1}{y_C} + \frac{1}{N_C - y_C},
$$
where $y_t$ are the number of events and $N_t$ are the total number of patients for treatment $t$.

A full list including alternative treatment effects and variance components available for binary outcome data is given in Table~\ref{tab:releff}.

\paragraph{Running oustandR()}
Having completed the set-up steps, we are now ready to proceed with performing the indirect treatment comparison with \texttt{outstandR()}. In each of the calls, we will explicitly set the random seed for reproducibility.

For the MAIC approach, we define the outcome and balancing formulas explicitly in a list. This explicitly separates the statistical intent of weighting versus regression:

\begin{verbatim}
> outstandR_maic <-
+   outstandR(ipd_trial = AC_IPD_binY_contX,
+             ald_trial = BC_ALD_binY_contX,
+             strategy = strategy_maic(
+               formula = list(outcome_model = formula("y ~ trt"),
+                              balance_model =  bal_form_contX),
+               family = binomial(link = "logit")),
+             seed = 12345)
\end{verbatim}

As detailed in Section~\ref{ssec:outstandr_function}, the \texttt{outstandR()} function returns a list object. To help with interpreting the results, \pkg{outstandR} includes a convenience \texttt{print} method.

\begin{verbatim}
> print(outstandR_maic)

Object of class 'outstandR' 
ITC algorithm: MAIC 
Model: binomial 
Scale: log_odds 
Common treatment: C 
Individual patient data study: A vs C 
Aggregate level data study: B vs C 
Confidence interval level: 0.95 

Contrasts:

# A tibble: 3 × 5
  Treatments Estimate Std.Error lower.0.95 upper.0.95
  <chr>         <dbl>     <dbl>      <dbl>      <dbl>
1 AB            0.616     0.251     -0.365     1.60  
2 AC           -0.828     0.149     -1.58     -0.0723
3 BC           -1.44      0.102     -2.07     -0.817 

Absolute:

# A tibble: 3 × 5
  Treatments Estimate Std.Error lower.0.95 upper.0.95
  <chr>         <dbl>     <dbl>      <dbl>      <dbl>
1 A             0.244   0.00216      0.153      0.335
2 B            -1.02    0.108       -1.67      -0.379
3 C             0.422   0.00548      0.277      0.567
\end{verbatim}

\begin{figure}[!ht]
    \centering
    \includegraphics[width=1.0\linewidth]{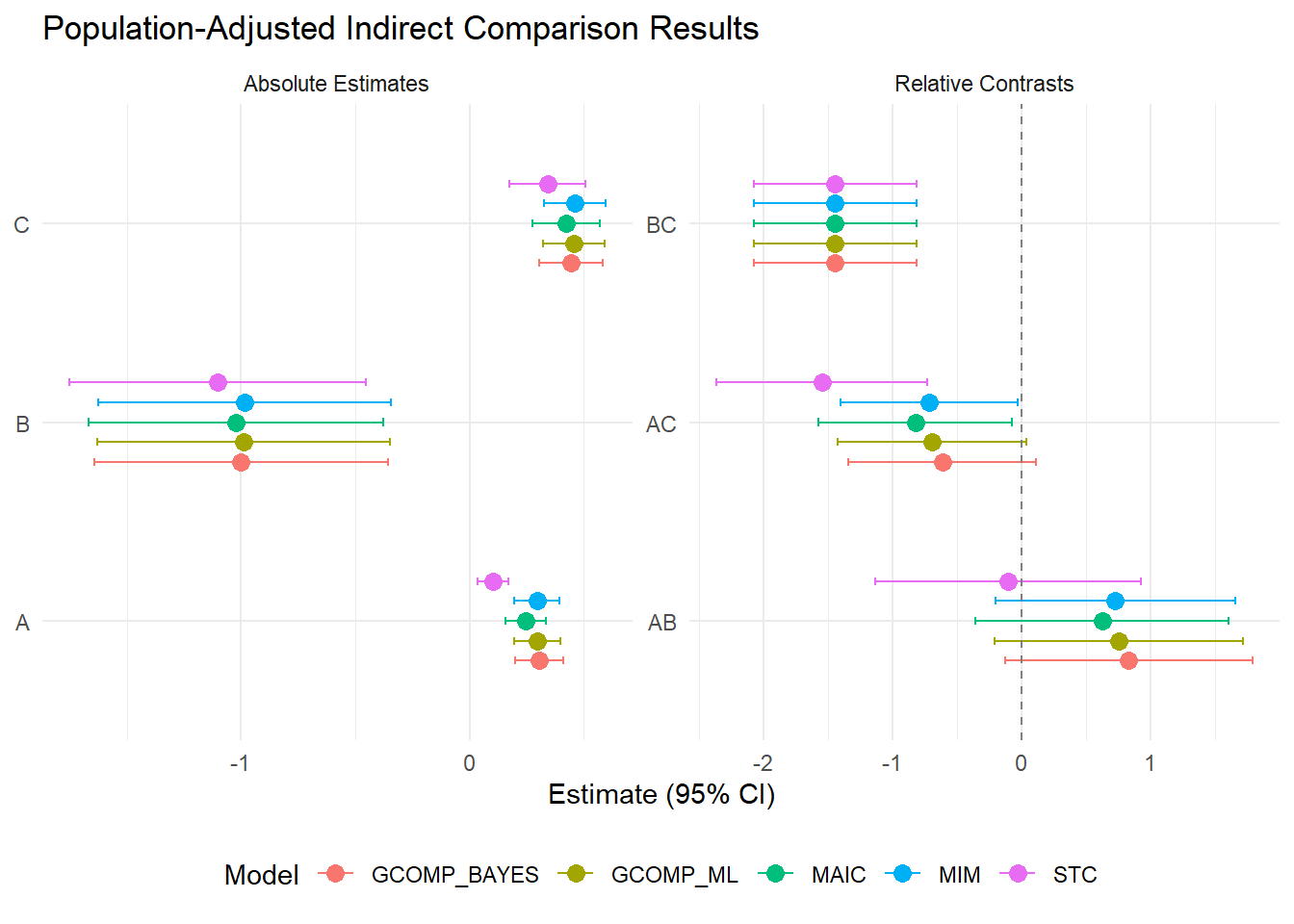}
    \caption{Forest plots of aboslute estimates and relative contrasts for all adjustment strategies.}
    \label{fig:summary-forest-plot}
\end{figure}

This contains the same information as the raw output but in a more presentable format. The first block of text provides modelling details, and the following tibbles show the contrast and absolute effects, respectively.
The STC, G-computation and MIM analyses are performed similarly but with the \texttt{outcome\_model} specified instead of the \texttt{balance\_model}.
%
%
%
Forest plots of the results across all adjustment strategies for the binary outcome are given in Figure~\ref{fig:summary-forest-plot} and generated by calling the following. 

\begin{verbatim}
> plot(outstandR_maic, outstandR_stc, outstandR_gcomp_ml, 
+      outstandR_gcomp_bayes, outstandR_mim)
\end{verbatim}

\subsubsection{Continuous outcome data}
Analogously to the first example with binary outcomes and continuous covariates, we now extend this for the case of continuous outcomes and a mixture of binary and continuous covariates. We deliberately include binary covariates in this tutorial to reflect the realities of evidence synthesis, where discrete effect modifiers are ubiquitous. However, as noted in Section~\ref{sec:synth-cohort}, applying a Gaussian copula to non-continuous marginals is a mathematical approximation. In practice, this may be preferable to omitting crucial discrete confounders entirely.

The same steps are followed as previously, first, reading in the data.

\begin{verbatim}
> data(AC_IPD_contY_mixedX) # A vs C individual-level data
> data(BC_ALD_contY_mixedX) # B vs C aggregate-level data
\end{verbatim}




The formats are the same as the previous binary data example but now in the ALD the binary-valued covariates have \texttt{statistic} entry \texttt{prop}. Covariates \texttt{X1} and \texttt{X3} are continuous and \texttt{X2} and \texttt{X4} are binary.

The formula for this example is also a little more complicated. We shall assume that some of the covariate can be both prognostic factors and effect modifiers, so the regression is

$$
y = \alpha + \beta_{01} X_1 + \beta_{02} X_2 + \beta_{03} X_3 + (\beta_{trt} + \beta_{11} X_1 + \beta_{12} X_2 + \beta_{14} X_4 ) \mathbb{I}(trt = 1) 
$$

This corresponds to the \proglang{R} formula given by
\begin{verbatim}
> lin_form_mixedX <- as.formula("y ~ X1 + X2 + X3 + trt + trt:(X1 + X2 + X4)")
\end{verbatim}

The balance formula includes the effect modifiers and is given by
\begin{verbatim}
> bal_form_mixedX <- as.formula("~ X1 + X2 + X4")
\end{verbatim}

\paragraph{Relative treatment effect}
Let $\hat{\mu}_B$ be the mean in the comparator arm and $\hat{\mu}_C$ be the mean in the reference arm. Then the mean treatment difference required in Equation~(\ref{eqn:paic}) is simply
$$
\hat{\Delta}_{BC} = \hat{\mu}_B - \hat{\mu}_C.
$$
For other scales on continuous data, the approach may involve transforming the means directly, although typical continuous outcome analyses focus on mean differences.

The variance of the relative treatment effect is
$$
\text{Var}(\Delta_{BC}) = \frac{s_B^2}{N_B} + \frac{s_C^2}{N_C}.
$$
where, for treatment arm $t$, $N_t$ and $s_t^2$ are the number of patients and the observed standard deviation, respectively.

\paragraph{Running outstandR()}
A difference from the previous binary outcome data example is the \texttt{family = gaussian(link = "identity")} argument indicating a normal linear regression. The full call to \texttt{outstandR()} is

\begin{verbatim}
> outstandR_maic <-
+   outstandR(ipd_trial = AC_IPD_contY_mixedX, 
+             ald_trial = BC_ALD_contY_mixedX,
+             strategy = strategy_maic(
+               formula = list(outcome_model = formula("y ~ trt"),
+                              balance_model = bal_form_mixedX),
+               family = gaussian(link = "identity")),
+             seed = 12345)
\end{verbatim}

Again, if we interrogate the output we see the same list format. The differences from the first example are that the model and scale attributes correspond to the input data used. That is, the default output contrast is the mean difference.

\begin{verbatim}
> str(outstandR_maic, max.level = 1)

List of 11
 $ results   :List of 2
 $ call      : language outstandR(ipd_trial = AC_IPD_contY_mixedX, __truncated__
 $ formula   :Class 'formula'  language y ~ X1 + X2 + X3 + trt + trt:(X1 + X2 + X4)
  .. ..- attr(*, ".Environment")=<environment: R_GlobalEnv> 
 $ CI        : num 0.95
 $ ref_trt   : chr "C"
 $ ipd_comp  : chr "A"
 $ ald_comp  : chr "B"
 $ scale     : chr "mean_difference"
 $ var_method: chr "sample"
 $ family    : chr "gaussian"
 $ model     :List of 3
 - attr(*, "class")= chr [1:2] "outstandR" "list

> print(outstandR_maic)

Object of class 'outstandR' 
ITC algorithm: MAIC 
Model: gaussian 
Scale: mean_difference 
Common treatment: C 
Individual patient data study: A vs C 
Aggregate level data study: B vs C 
Confidence interval level: 0.95 

# A tibble: 3 × 5
  Treatments Estimate Std.Error lower.0.95 upper.0.95
  <chr>         <dbl>     <dbl>      <dbl>      <dbl>
1 AB           -0.181    0.132      -0.892      0.529
2 AC           -1.11     0.100      -1.73      -0.487
3 BC           -0.926    0.0311     -1.27      -0.581

Absolute:

# A tibble: 3 × 5
  Treatments Estimate Std.Error lower.0.95 upper.0.95
  <chr>         <dbl>     <dbl>      <dbl>      <dbl>
1 A            -0.579    0.0304    -0.921      -0.238
2 B            -0.398    0.0925    -0.994       0.198
3 C             0.528    0.0614     0.0425      1.01 
\end{verbatim}

The corresponding calls to \texttt{outstandR()} for the other analyses follow similarly.

\subsubsection{Count outcome data}
First, read in and view the two data sets.

\begin{verbatim}
> data(AC_IPD_countY_contX) # A vs C individual-level data
> data(BC_ALD_countY_contX) # B vs C aggregate-level data
\end{verbatim}





The formats are the same as the binary outcome data example.
Let $\lambda$ be the outcome event rate, then the equation for the outcome regression for the count data is
$$
\log(\lambda) = \alpha + \beta_{01} X_{1} + \beta_{02} X_{2} + (\beta_{trt} + \beta_{13} X_{3} + \beta_{14} X_{4}) \mathbb{I}(trt = 1).
$$

Recalling that the log link is defined separately in the \texttt{outstandR()} function call, we shall use the same linear formulae as for the binary outcome example, \texttt{lin\_form\_contX} and \texttt{bal\_form\_contX}.

\paragraph{Relative treatment effects}
The default relative effect with the log link is the log relative risk, with mean
$$
\hat{\Delta}_{BC} = \ln(\hat{\mu}_B) - \ln(\hat{\mu}_C)
$$ 
and variance
$$
\text{Var}(\Delta_{BC}) = \frac{1}{N_B \hat{\mu}_B} + \frac{1}{N_C \hat{\mu}_C}
$$ 

Table~\ref{tab:releff} gives the full list of potential relative effects for count outcome data.

\paragraph{Running outstandR()}
A difference from the previous outcome data examples is the \texttt{family = poisson(link = "log")} argument indicating a log linear regression. The full call to \texttt{outstandR()} is

\begin{verbatim}
> outstandR_maic <-
+   outstandR(ipd_trial = AC_IPD_countY_contX,
+             ald_trial = BC_ALD_countY_contX,
+             strategy = strategy_maic(
+               formula = list(outcome_model = formula("y ~ trt"),
+                              balance_model = bal_form_contX),
+               family = poisson(link = "log")),
+             seed = 12345)
\end{verbatim}

The differences from the first example are that the model and scale attributes correspond to the input data used. That is, the default output contrast is the mean difference.

\begin{verbatim}
> print(outstandR_maic)

Object of class 'outstandR' 
ITC algorithm: MAIC 
Model: poisson 
Scale: log_relative_risk 
Common treatment: C 
Individual patient data study: A vs C 
Aggregate level data study: B vs C 
Confidence interval level: 0.95 

Contrasts:

# A tibble: 3 × 5
  Treatments Estimate Std.Error lower.0.95 upper.0.95
  <chr>         <dbl>     <dbl>      <dbl>      <dbl>
1 AB           0.0973    0.0727     -0.431      0.626
2 AC          -1.16      0.0471     -1.59      -0.736
3 BC          -1.26      0.0256     -1.57      -0.945

Absolute:

# A tibble: 2 × 5
  Treatments Estimate Std.Error lower.0.95 upper.0.95
  <chr>         <dbl>     <dbl>      <dbl>      <dbl>
1 A            0.403    0.00384      0.281      0.524
2 B            0.0313   0.0581      -0.441      0.504
3 C            1.29     0.0325       0.936      1.64 
\end{verbatim}
The corresponding calls to \texttt{outstandR()} for the other analyses follow in a similar way.

\subsection{User-specified outcome scales}
By default, \pkg{outstandR} reports treatment effects on the scale corresponding to the link function of the outcome regression. For instance, a logistic regression specified with \texttt{family = binomial(link = "logit")} naturally yields estimates in log-odds ratios, while a linear regression yields mean differences. However, health economic models often require parameters on specific scales, such as risk differences (RD) or relative risks (RR), to facilitate interpretation or align with decision-analytic model inputs.

\begin{table}[!ht]
    \centering
    \renewcommand{\arraystretch}{2.2}
    \begin{tabular}{@{} l l c c @{}}
        \toprule
        \textbf{Outcome Data} & \textbf{Relative Effect} & \textbf{Mean Difference} & \textbf{Treatment Variance} \\
        \midrule
        Binary 
          & Log Odds Ratio$^\dagger$ & $\text{logit}(p_B) - \text{logit}(p_C)$ & $\frac{1}{y_t} + \frac{1}{N_t - y_t}$ \\
          & Risk Difference & $p_B - p_C$ & $\frac{p_t(1 - p_t)}{N_t}$ \\
          & Probit Difference & $\text{probit}(p_B) - \text{probit}(p_C)$ & $\frac{1}{y_t} + \frac{1}{N_t - y_t}$ \\
          & Log RR (cloglog) & $\ln(-\ln(1 - p_B)) - \ln(-\ln(1 - p_C))$ & $\frac{1}{y_t} - \frac{1}{N_t}$ \\
          & Log RR (Log Link) & $\ln(p_B) - \ln(p_C)$ & $\frac{1}{y_t} - \frac{1}{N_t}$ \\
        Count 
          & Log Relative Risk$^\dagger$ & $\ln(\hat{\mu}_B) - \ln(\hat{\mu}_C)$ & $\frac{1}{N_t \hat{\mu}_t}$ \\
          & Rate Difference & $\hat{\mu}_B - \hat{\mu}_C$ & $\frac{\hat{\mu}_t}{N_t}$ \\
        Continuous 
          & Mean Difference$^\dagger$ & $\hat{\mu}_B - \hat{\mu}_C$ & $\frac{s_t^2}{N_t}$ \\
        \bottomrule
    \end{tabular}
    \caption{Relative treatment effect statistics for different outcome data types. For binary data, $y_t$ = events, $N_t$ = total patients. For count data, $\hat{\mu}_t$ = mean rate/count, $N_t$ = patients. $^\dagger$default in \pkg{outstandR}.}
        \label{tab:releff}
\end{table}

Table~\ref{tab:releff} gives all of the available relative effects for each outcome data type. The \texttt{scale} argument in \texttt{outstandR()} allows users to request these alternative summary statistics directly. 
When converting between relative measures (e.g., odds ratios) and absolute measures (e.g., risk differences), the package automatically estimates the necessary baseline risk ($P_0$) from the reference treatment arm in the IPD study.

For log-odds and log-relative risk on continuous data, the \proglang{R} code implies a transformation of the mean before calculating the variance. For \code{log\_odds}, it uses $\frac{\pi^2}{3 N_t}$, which is the variance of a logit-transformed mean when the underlying data are assumed to be normal and mapped to $(0,1)$ scale, and then transformed to logit (often used for proportions derived from continuous data or for scores that can be thought of as probabilities).

For example, to obtain a risk difference instead of the default log-odds ratio for the binary outcome analysis (previously shown in Section~\ref{sec:examples}), we specify \code{scale = "risk\_difference"}.\\
\\
\begin{minipage}{\linewidth}
\begin{verbatim}

> outstandR_maic_rd <- outstandR(
+   ipd_trial = AC_IPD_binY_contX, 
+   ald_trial = BC_ALD_binY_contX,
+   strategy = strategy_maic(
+     formula = list(outcome_model = formula("y ~ trt"),
+                    balance_model = bal_form_contX),
+     family = binomial(link = "logit")),
+   seed = 12345,
+   scale = "risk_difference"  # instead of default log-odds ratio
+ )

\end{verbatim}
\end{minipage}

The output now displays the contrast estimates on the probability scale:

\begin{verbatim}
> print(outstandR_maic_rd)

Object of class 'outstandR' 
ITC algorithm: MAIC 
Model: binomial 
Scale: risk_difference 
Common treatment: C 
Individual patient data study: A vs C 
Aggregate level data study: B vs C 
Confidence interval level: 0.95 

Contrasts:

# A tibble: 3 × 5
  Treatments Estimate Std.Error lower.0.95 upper.0.95
  <chr>         <dbl>     <dbl>      <dbl>      <dbl>
1 AB            0.164   0.442       -1.14      1.47  
2 AC           -0.177   0.00675     -0.338    -0.0161
3 BC           -0.341   0.436       -1.63      0.952 

Absolute:

# A tibble: 3 × 5
  Treatments Estimate Std.Error lower.0.95 upper.0.95
  <chr>         <dbl>     <dbl>      <dbl>      <dbl>
1 A            0.244    0.00197      0.157      0.331
2 B            0.0801   0.441       -1.22       1.38 
3 C            0.421    0.00525      0.279      0.563
\end{verbatim}

The package determines the validity of a conversion path based on the underlying data distribution; for instance, requesting a \code{mean\_difference} for a binary outcome model will raise a validation error.

\subsection{Robust variance estimation}
Standard errors in population-adjusted analyses can be sensitive to model misspecification. Users can request robust (sandwich) standard errors (see Section~\ref{sec:variance_methods}) by setting the \code{var\_method} argument to \code{"sandwich"}. 

The following example demonstrates how to compare the naive model-based standard error with the robust sandwich estimate for a maximum likelihood G-computation analysis:

\begin{verbatim}
> # standard G-computation ML analysis (model-based variance)
> stc_naive <- outstandR(
+    ipd_trial = AC_IPD_binY_contX,
+    ald_trial = BC_ALD_binY_contX,
+    strategy = strategy_gcomp_ml(
+      formula = list(outcome_model = lin_form_contX),
+      family = binomial()),
+    seed = 12345,
+    var_method = "sample") # default

> # G-computation ML with robust variance (sandwich estimator)
> stc_robust <- outstandR(
+    ipd_trial = AC_IPD_binY_contX,
+    ald_trial = BC_ALD_binY_contX,
+    strategy = strategy_gcomp_ml(
+      formula = list(oucome_model = lin_form_contX),
+      family = binomial()),
+    seed = 12345,
+    var_method = "sandwich")

> # compare standard errors
> stc_naive$contrasts$var
> stc_robust$contrasts$var

       AB        AC        BC 
0.2727807 0.1705496 0.1022311 

       AB        AC        BC 
0.2266784 0.1244473 0.1022311 
\end{verbatim}

This option allows analysts to assess the stability of their conclusions without requiring manual implementation of the delta method or bootstrapping.

\subsection{User-defined covariate distributions in G-computation}
By default, the simulation step in G-computation assumes that the covariates in the target population follow a multivariate normal distribution, parameterized by the means and standard deviations reported in the ALD (corresponding to $F_k(\cdot; \theta_k)$ in Equation~(\ref{eqn:marginal-transformation})). However, this assumption may be inappropriate. For example, age or costs may be skewed and strictly positive, while binary variables, such as sex, are bounded.

\pkg{outstandR} allows users to specify alternative marginal distributions for specific covariates via the \code{marginal\_distns} argument.

A key feature of this implementation is that the package automatically parameterizes these distributions using the summary statistics available in the ALD. For example, if a user specifies a Gamma distribution for \code{age}, \pkg{outstandR} will automatically retrieve the reported mean ($\mu$) and standard deviation ($\sigma$) from the ALD data frame and convert them into the required shape ($\alpha$) and rate ($\beta$) parameters using the method of moments:
\begin{equation*}
\alpha = (\mu / \sigma)^2, \quad \beta = \mu / \sigma^2.
\end{equation*}
This is more robust and streamlines the workflow by removing the need for users to perform these statistical transformations manually.

First, define the target distributions. We specify a Gamma distribution with “gamma” for \texttt{X1} and a Binomial with “binom” for \texttt{X2}. The other covariates not listed (\texttt{X3}, \texttt{X4}) default to a normal distribution, i.e., “norm”.

\begin{verbatim}
> # define distributions
> custom_distns <- c(X1 = "gamma", X2 = "binom")

> # run G-computation ML
> outstandR_custom <- outstandR(
+    ipd_trial = AC_IPD_contY_mixedX,
+    ald_trial = BC_ALD_contY_mixedX,
+    strategy = strategy_gcomp_ml(
+      formula = list(outcome_model = lin_form_mixedX),
+      family = gaussian(link = "identity"),
+      marginal_distns = custom_distns,
+      N = 1000),
+    seed = 12345)

> print(outstandR_custom)

Object of class 'outstandR' 
ITC algorithm: GCOMP_ML 
Model: gaussian 
Scale: mean_difference 
Common treatment: C 
Individual patient data study: A vs C 
Aggregate level data study: B vs C 
Confidence interval level: 0.95 

Contrasts:

# A tibble: 3 × 5
  Treatments Estimate Std.Error lower.0.95 upper.0.95
  <chr>         <dbl>     <dbl>      <dbl>      <dbl>
1 AB           -0.285    0.0892     -0.871      0.300
2 AC           -1.21     0.0581     -1.68      -0.739
3 BC           -0.926    0.0311     -1.27      -0.581

Absolute:

# A tibble: 3 × 5
  Treatments Estimate Std.Error lower.0.95 upper.0.95
  <chr>         <dbl>     <dbl>      <dbl>      <dbl>
1 A            -0.514    0.0170     -0.770     -0.259
2 B            -0.229    0.0765     -0.772      0.313
3 C             0.697    0.0454      0.279      1.11 
\end{verbatim}

This approach ensures that the synthetic cohort respects the natural constraints of the data (e.g., non-negative age) while strictly adhering to the evidence reported in the aggregate summary.

\paragraph{Manual parameter specification}
There are scenarios where manual specification via the \code{marginal\_params} argument is also necessary. 
One use case is as a sensitivity analysis, where a researcher wishes to assess the robustness of the relative treatment effect estimates to deviations in the target population's characteristics of the IPD trial. For instance, if the ALD reports a mean age of 65, a researcher might wish to simulate a target population with a mean age of 70 to test if the conclusions hold for an older cohort. Additionally, manual specification is required for distributions where the method of moments conversion is not natively supported by the package (e.g., specific parameterizations of the Weibull distribution).
The following code illustrates how to manually override the ALD statistics to simulate an older population.

\begin{verbatim}
> # manually specify parameters for 'age' (mean=70, sd=10.5)
> # convert mean=70/sd=10.5 to shape/rate manually 
> # or pass them if we already know the specific shape/rate
> target_mean <- 70
> target_sd <- 10.5

> sens_params <- list(
+   X1 = list(shape = (target_mean/target_sd)^2, 
+             rate = target_mean/(target_sd^2)))

> # run G-computation with manual override
> outstandR_sens <- outstandR(
+    ipd_trial = AC_IPD_contY_mixedX,
+    ald_trial = BC_ALD_contY_mixedX,
+    strategy = strategy_gcomp_ml(
+      formula = list(outcome_model = lin_form_mixedX),
+      family = gaussian(link = "identity"),
+      marginal_distns = new_distns,
+      marginal_params = new_params, # overrides ALD for X1
+      N = 1000),
+    seed = 12345)

> print(outstandR_sens)

Object of class 'outstandR' 
ITC algorithm: GCOMP_ML 
Model: gaussian 
Scale: mean_difference 
Common treatment: C 
Individual patient data study: A vs C 
Aggregate level data study: B vs C 
Confidence interval level: 0.95 

Contrasts:

# A tibble: 3 × 5
  Treatments Estimate Std.Error lower.0.95 upper.0.95
  <chr>         <dbl>     <dbl>      <dbl>      <dbl>
1 AB          -22.1   1160.         -88.9      44.6  
2 AC          -23.1   1160.         -89.8      43.7  
3 BC           -0.926    0.0311      -1.27     -0.581

Absolute:

# A tibble: 3 × 5
  Treatments Estimate Std.Error lower.0.95 upper.0.95
  <chr>         <dbl>     <dbl>      <dbl>      <dbl>
1 A              62.4      271.       30.2       94.7
2 B              84.6      914.       25.3      144. 
3 C              85.5      914.       26.2      145
\end{verbatim}

\subsection{Model diagnostics}
While \texttt{outstandR()} focuses on estimation, assessing model fit is crucial. The returned S3 object exposes the underlying statistical artifacts in the \texttt{analysis\_data} slot, allowing users to leverage external diagnostic tools.

For example, for an MAIC analysis, we can inspect the distribution of weights to check for outliers or near-zero effective sample sizes.

\begin{figure}[!ht]
    \centering
    \includegraphics[width=0.8\linewidth]{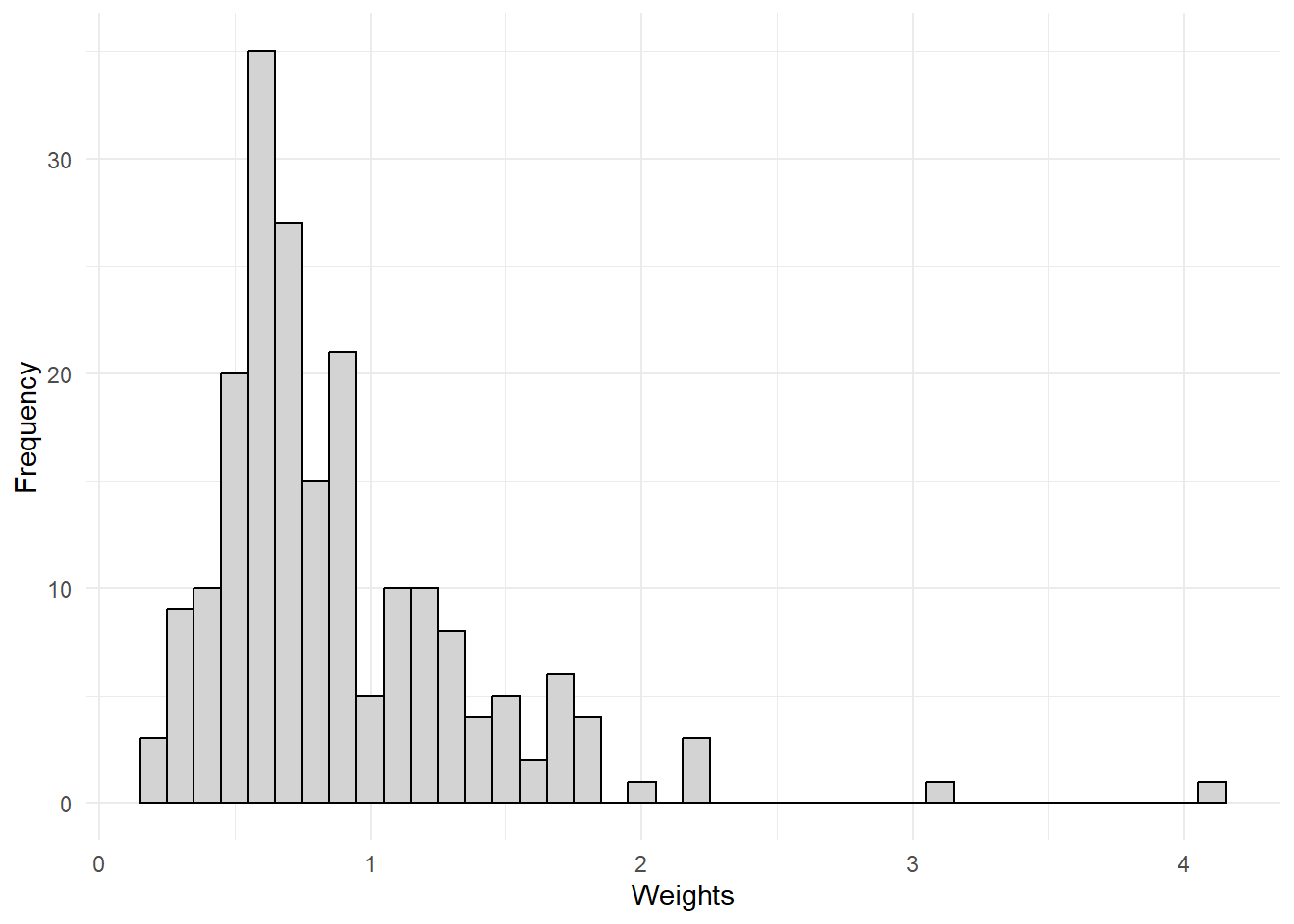}
    \caption{Histogram of MAIC analysis distribution of weights.}
    \label{fig:maic-diagnostic}
\end{figure}

\begin{verbatim}
> # check Effective Sample Size
> print(outstandR_maic$model$ESS) 
    ESS 
151.053 

> # histogram of weights
> library(ggplot2)

> data.frame(weights = outstandR_maic$model$weights) |> 
+   ggplot(aes(x = weights)) +
+   geom_histogram(binwidth = 0.1, color = "black", fill = "lightgray") +
+   labs(x = "Weights", y = "Frequency") +
+   theme_minimal()
\end{verbatim}
Figure~\ref{fig:maic-diagnostic} shows the histogram of MAIC weights.
Similarly, for G-computation, users can extract the fitted model to check residuals or, in the Bayesian case, inspect MCMC convergence using, for example, the \pkg{bayesplot} package.

\begin{figure}[!ht]
    \centering
    \includegraphics[width=0.8\linewidth]{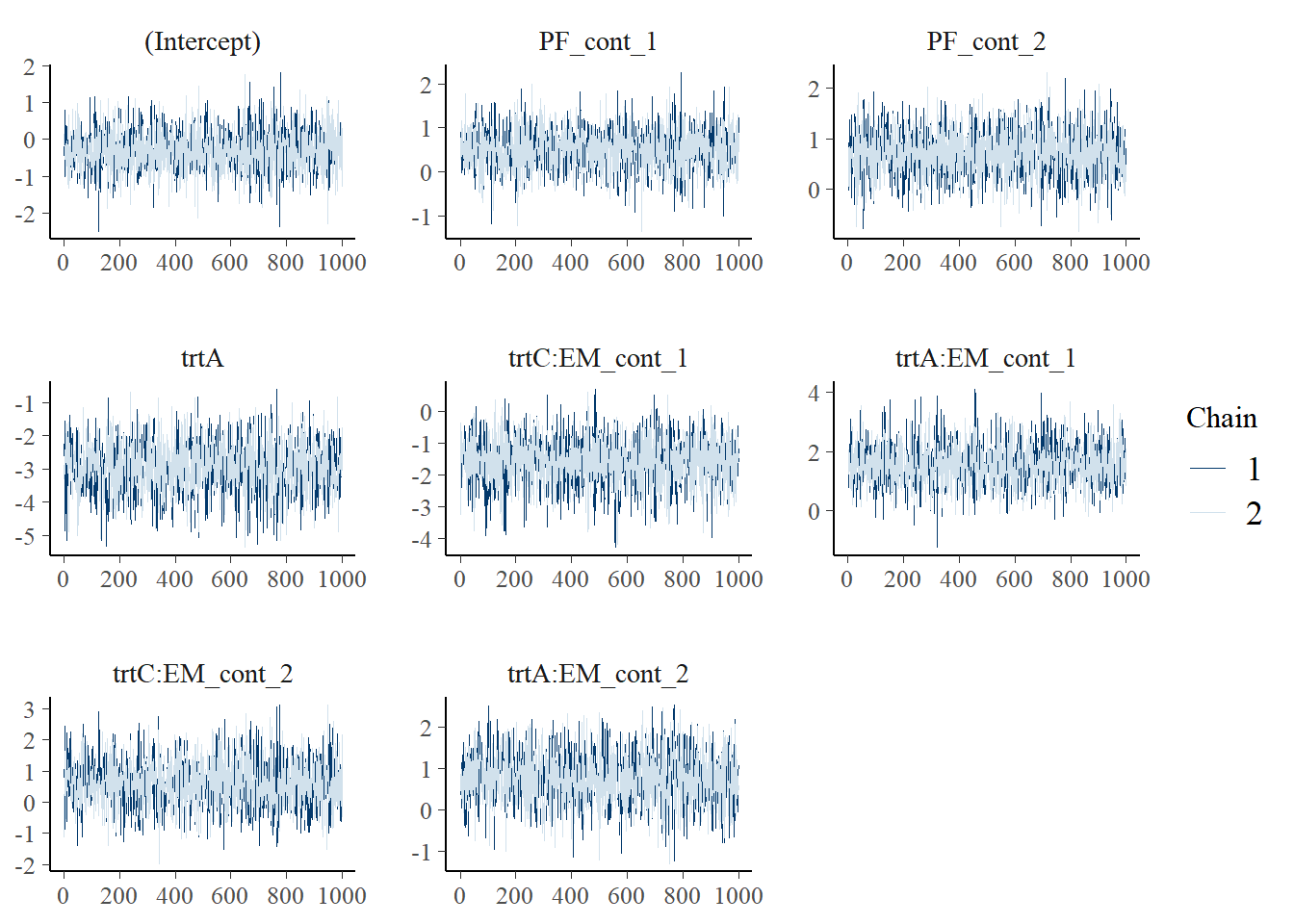}
    \caption{Bayesian G-computation MCMC trace plots.}
    \label{fig:gcomp-diagnostic}
\end{figure}

\begin{verbatim}
> # extract stanfit object
> stan_obj <- outstandR_gcomp_bayes$model$fit 

> # use external package for diagnostics
> library(bayesplot)
> mcmc_trace(stan_obj)
\end{verbatim}
Figure~\ref{fig:gcomp-diagnostic} shows the MCMC trace plots.

\section{Conclusion}
\label{sec:conclusion}

The \pkg{outstandR} package offers a comprehensive and flexible toolset to address indirect treatment comparisons, especially in the challenging context of limited individual patient data. Its key strengths include:
\begin{itemize}
    \item Versatility: Support for multiple standardisation methods (MAIC, STC, G-computation, MIM) allows users to choose the most appropriate approach for their data and assumptions.
    \item Uncertainty quantification: The Bayesian G-computation approach provides a robust framework for propagating and quantifying uncertainty.
    \item Practical relevance: Directly addresses a critical need in health economics and evidence synthesis to compare treatments when direct trial data is unavailable.
\end{itemize}
While powerful, \pkg{outstandR} relies on the correct specification of outcome models and the availability of sufficient aggregate-level data for reliable estimation. For example, matching on higher-order moments with the MAIC approach may reduce bias by matching joint distributions, it drastically increases the parameters the optimizer must solve for - an example of the bias-variance trade-off - and commonly results in more extreme weights and a lower ESS.
These are active areas of research. Future work will include expanding the range of supported outcome models and exploring more advanced covariate simulation techniques.

The package’s modular S3 architecture and explicit two-part formula design (\texttt{outcome\_model} and \texttt{balance\_model}) were designed specifically to accommodate future methodological extensions. One such approach is that of so-called \textit{doubly robust} estimators \citep{campbell2025}. These methods combine regression and weighting to allow for ``two chances of obtaining a correct model''; \pkg{outstandR}'s new formula architecture natively captures the necessary inputs for these upcoming methods.

We also intend to implement functionality and workflows for the case when a common comparator is absent using single-arm studies called \textit{unanchored comparisons}. We will allow for the synthesis of multiple ALD trials into a wider network of evidence. Finally, while \pkg{outstandR} currently intercepts \texttt{Surv} objects with an informative error, comprehensive support for time-to-event (survival) data is officially scheduled for the upcoming release, leveraging the established \pkg{survival} package \citep{survivalpkg}.

\bibliography{references} 

\end{document}